\documentclass[aps,preprint]{revtex4-1}

\pdfoutput=1

\usepackage[T1]{fontenc}
\usepackage{lmodern}
\usepackage{hyperref}
\usepackage{amsmath}
\usepackage{amssymb}
\usepackage{graphicx}
\usepackage{color}
\usepackage{amssymb}
\usepackage{amsthm}
\usepackage{epstopdf} 
\usepackage{dcolumn}
\usepackage{amsfonts}
\usepackage{bm}

\usepackage[normalem]{ulem}
\newcommand{\gvm}{\,\mathrm{GV/m}}
\newcommand{\cc}{\,\mathrm{cm^{-3}}}
\newcommand{\wcm}{\,\mathrm{W/cm}^2}
\newcommand{\mic}{\,\mu\mathrm{m}}

\newcommand{\MeV}{\,\mathrm{MeV}}

\newcommand{\fs}{\,\mathrm{fs}}
\newcommand{\pC}{\,\mathrm{pC}}

\newcommand{\mrad}{\,\mathrm{mrad}}

\begin{document}


\title{Relativistic electron beams driven by kHz single-cycle light pulses
}

\author{$^{1}$D. Gu\'enot, $^{1}$D. Gustas, $^1$A. Vernier, $^1$B. Beaurepaire, $^1$F. B\"ohle, $^1$M. Bocoum, $^1$M. Lozano, $^1$A. Jullien, $^1$R. Lopez-Martens, $^1$A. Lifschitz and $^1$J. Faure$^*$}

\affiliation{$^1$LOA, ENSTA Paristech, CNRS, Ecole Polytechnique, Universit\'e Paris-Saclay, Palaiseau, France} 

\begin{abstract}


Laser-plasma acceleration \cite{taji79,esar09} is an emerging technique for accelerating electrons to high energies over very short distances. The accelerated electron bunches have femtosecond duration \cite{lund11,buck11}, making them particularly relevant for applications such as ultrafast imaging \cite{miller14} or femtosecond X-ray generation \cite{corde13,taph12}. Current laser-plasma accelerators are typically driven by Joule-class laser systems that have two main drawbacks: their relatively large scale and their low repetition-rate, with a few shots per second at best. The accelerated electron beams have energies ranging from 100~MeV \cite{faur04,gedd04,mang04} to multi-GeV \cite{wang13,leem14}, however a MeV electron source would be more suited to many societal and scientific applications.  Here, we demonstrate a compact and reliable laser-plasma accelerator producing high-quality few-MeV electron beams at kilohertz repetition rate. This breakthrough was made possible by using near-single-cycle light pulses, which lowered the required laser energy for driving the accelerator by three orders of magnitude, thus enabling high repetition-rate operation and dramatic downsizing of the laser system. The measured electron bunches are collimated, with an energy distribution that peaks at 5 MeV and contains up to 1 pC of charge. Numerical simulations reproduce all experimental features and indicate that the electron bunches are only $\sim 1$~fs long. We anticipate that the advent of these kHz femtosecond relativistic electron sources will pave the way to wide-impact applications, such as ultrafast electron diffraction in materials \cite{zewa06,scia11} with an unprecedented sub-10 fs resolution \cite{faure16}.

\end{abstract}

\maketitle
In a laser-plasma accelerator, a laser pulse is focused to ultra-high intensity in an underdense plasma. The laser ponderomotive force sets up a charge separation in the plasma by displacing electrons, resulting in the excitation of a large-amplitude plasma wave, also called a wakefield. The wakefield carries enormous electric fields, in excess of $100\gvm$ \cite{malk02}, that are well adapted for accelerating electrons to relativistic energies over short distances, typically less than a millimeter. The accelerated electron beams have femtosecond duration and are intrinsically synchronized to the laser pulse, which could lift the temporal resolution bottleneck in various experimental situations. For example, in ultrafast electron diffraction, the temporal resolution is currently limited to more than 100~fs, but it could be improved to sub-10 fs using laser driven electrons \cite{faure16}. Thus, laser-plasma accelerators in the MeV range could find numerous applications with unprecedented time resolution, provided they operate reliably and at high repetition-rate. Indeed, in addition to temporal resolution, ultrafast imaging and diffraction also require statistics and a high signal-to-noise ratio \cite{miller14,scia11} that can only be reached with a reliable and high repetition-rate electron source.

 In this letter, we demonstrate reliable operation of a laser-plasma accelerator delivering 5~MeV electrons at kHz repetition-rate. This breakthrough was made  possible by the original use of  a multi-mJ laser system  delivering near-single-cycle laser pulses of 3.4~fs duration \cite{boeh14,jull14} (see Methods). In addition, this work demonstrates the scalability of laser wakefield acceleration to the mJ energy level, enabling the use of box-size and commercial laser systems for driving laser-plasma accelerators.

It is well established that the blow-out, or bubble regime of laser-plasma acceleration \cite{pukh02,lu06,faur04,gedd04,mang04} leads to the production of  high-quality relativistic electron beams with narrow energy spreads and small divergence. In this regime, the laser pulse is transversely and longitudinally resonant with the plasma wavelength $\lambda_p$, i.e. its longitudinal and transverse sizes are comparable to $\lambda_p$: $c\tau \approx w_0 \approx \lambda_p/2$, where $c$ is the speed of light, $\tau$ the pulse duration, and $w_0$ the laser beam waist. When this condition is met at high intensity, $I\in10^{18}-10^{19}\wcm$, this results in the excitation of a very nonlinear wakefield, which takes the form of successive ion cavities surrounded by thin sheets of electrons \cite{lu06,lu07}. Electrons are injected in the accelerating field at the back of the ion cavity, thereby forming a bunch with femtosecond duration \cite{lund11}. In addition, the transverse fields of the cavity are focusing, which leads to a low divergence electron beam.

While this regime is routinely achieved in experiments using Joule-class laser systems with 30~fs laser pulses, scaling down to mJ-level, kHz laser systems constitutes a formidable challenge because of the scaling law of the bubble regime:
 $$
 E_L\propto \tau^3\propto\lambda_p^3,
 $$
where $E_L$ is the laser energy. The reduction of $E_L$ from J to mJ, a factor of 1000, must be accompanied by a reduction in both plasma wavelength and pulse duration by a factor of 10. The required pulse duration must drop from the standard $\sim$ 30~fs to only $\sim 3$~fs, i.e. the pulse should basically contain a single light cycle. It follows that the plasma wavelength must also be very small ($\lambda_p\simeq 2\mic$), corresponding to a high electron plasma density of $n_e\simeq 2.5\times10^{20}\cc$, which, according to scaling laws \cite{lu07} leads to acceleration of electrons in the 10 MeV range.
Previous attempts to accelerate electrons with mJ-class lasers \cite{he13,beau15,he13b,he16,goers15} did not reach the blow-out regime because the pulse duration was too long ($>20$~fs). Therefore, these initial experiments resulted in non relativistic 100~keV electrons beams \cite{he13,beau15,he13b} and did not yield high quality beams with femtosecond duration \cite{goers15}.   

\begin{figure*}[t!]
\centerline{\includegraphics[width=1\textwidth]{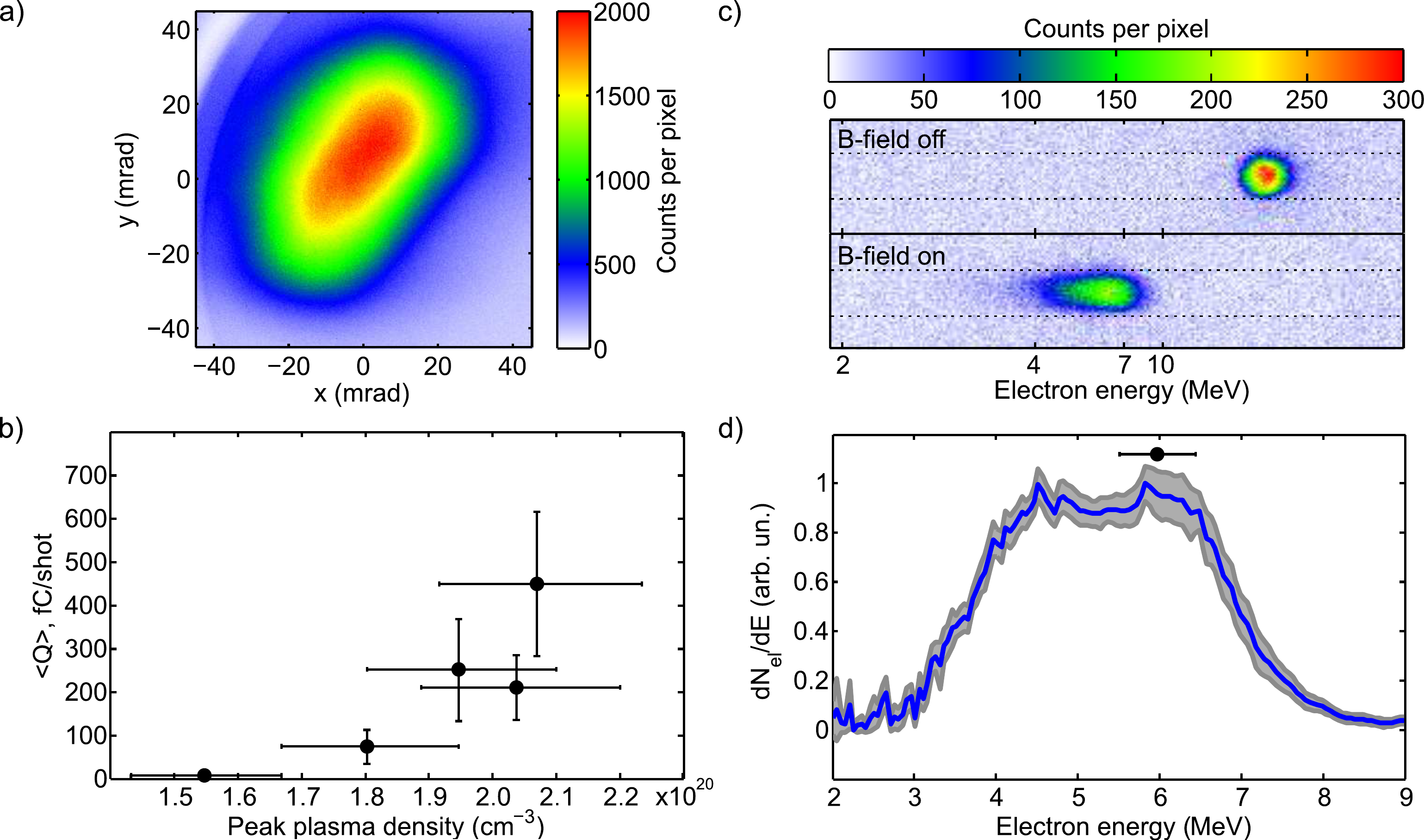}}
\caption{\textbf{Measurements of the kHz electron beam}. Panel a): Typical electron beam profile obtained by integrating over $500$ shots. The total beam charge is estimated to be $147$~fC/shot for this particular case. Panel b): Dependence of the beam charge as a function of the plasma density (the density was changed by varying the height of the gas jet). Vertical error bars represent the r.m.s. fluctuations whereas the horizontal error bars represent the uncertainty over the electron density. Panel c): Electron beam filtered by a $500\mic$ pinhole with and without magnetic field. The deviation of the electron spot by the magnetic field indicates acceleration to multi-MeV energies. Panel d): Electron spectrum; the grey area represents the standard deviation over $20$ spectra (each spectrum was obtained by accumulating over $1000$ laser shots). The horizontal error bar represents the spectrometer resolution at about $6\MeV$.}\label{fig1}
\end{figure*}

In the experiment, plasma waves are resonantly excited by focusing the $2.1\,\mathrm{mJ}$, $3.4\,\mathrm{fs}$ laser pulses into a continuously flowing, $100\,\mu\mathrm{m}$ diameter Nitrogen gas jet. The vacuum laser intensity is estimated at $\sim 3\times10^{18}\wcm$, allowing the pulse to ionize Nitrogen five times, therefore providing a background electron plasma density of $n_e\approx 1-2\times10^{20}\cc$. The electron spatial and spectral distributions are measured using standard diagnostics (see Methods). Figure~\ref{fig1} shows the characteristics of a typical electron beam observed in our experiment. As seen in Fig.~\ref{fig1}\,a), the beam has a rather small divergence of $\sim 45\,$mrad Full Width Half Maximum (FWHM). The beam pointing stability is high, with fluctuations amounting to a small fraction of the beam divergence,  typically a few mrad. Fig.~\ref{fig1}\,b) shows that the beam charge depends on the electron density, which can be increased by sending the laser closer to the nozzle. While an injected beam starts to appear for $n_e>1.5\times10^{20}\cc$, the charge goes up to $\sim0.5\,$pC/shot when the density approaches $2\times10^{20}\cc$, with the occasional observation of beams reaching 1 pC/shot. The rather large error bars in the figure show that the charge fluctuations are typically $30\%$~r.m.s., indicating that we operate close to the injection threshold, as discussed below.

The electron energy distribution is obtained by deflecting the electrons using an insertable pair of permanent magnets, as illustrated by the raw data presented in Fig.~\ref{fig1}\,c). A typical energy distribution is shown in Fig.~\ref{fig1}\,d). The electron spectrum peaks at about $5\MeV$, with an energy spread of $3\MeV$. The width of the grey line represents the r.m.s. fluctuations of the energy distribution showing that the acceleration process is very stable. 

The measured electrons beams, with their small divergence and peaked energy distribution, show typical features of acceleration in nonlinear, bubble-like wakefields. However, while these results confirm the scalability of laser-plasma acceleration, the use of single-cycle laser pulses reveals new physical effects that are usually insignificant at longer durations. For example, as the laser pulse is nearly composed of a single light cycle, the carrier envelope phase should have an effect on injection and acceleration \cite{lifs12}. In addition, such ultra-short pulses have an ultra-broad spectral bandwidth, spanning over an entire octave. Therefore, dispersion effects cannot be neglected during propagation in the plasma. In the linear limit, the pulse duration in the plasma evolves as $\tau(z)=\tau_0\sqrt{1+z^2/L_{disp}^2}$, where $\tau_0$ is the r.m.s. Fourier transform limited duration and $L_{disp}$ is the dispersion length scale in the plasma \cite{beau14}:
\begin{eqnarray}
L_{disp} \simeq 4\pi c^2\tau_0^2\frac{\lambda_p^2}{\lambda_0^3},
\label{eq:dispersion length}
\end{eqnarray}
For our parameters, $L_{disp}\simeq 20\mic$, indicating that the negative dispersion of the plasma causes rapid stretching of the single-cycle laser pulse. Therefore, in the experiment, we attempted to compensate this plasma dispersion by adding a small positive chirp to the laser pulse. Fig.~\ref{fig2}\,a) shows that the electron beam charge is maximum when a small positive chirp of $+8\,\mathrm{fs}^2$ is added to the laser pulse. This result was reproduced on multiple experimental runs, with the optimal chirp varying between $+4\,\mathrm{fs}^2$ and $+8\,\mathrm{fs}^2$. On the contrary, a negative chirp causes a decrease of the injected charge. Similarly, Fig.~\ref{fig2}\,b) shows that beam energy also increases when using a small positive chirp. This calls for a straightforward interpretation: a positive chirp compensates the negative plasma dispersion, allowing the laser pulse to reach higher intensities within the plasma. Consequently, the wakefield amplitude is higher, leading to higher injected charge and higher energy.

 Particle-In-Cell simulations reproduce our experimental results and confirm our interpretation. The simulations were run using the experimental parameters and a positive chirp of $+4\,\mathrm{fs}^2$ was added to the pulse (see Methods). Figure \ref{fig3}\,a) shows the evolution of the laser intensity during propagation in the plasma. The simulation shows that the initially chirped pulse compresses and self-focuses as it propagates. It reaches a high
intensity of $I\simeq 5.5\times 10^{18}\wcm$ around the middle of the plasma where the density is resonant with the laser pulse. At this point, the laser pulse is able to excite a high-amplitude wakefield. In addition, at this high intensity, $N^{5+}$ is ionized through tunnel ionization, triggering electron injection into the wakefield \cite{mcgu10,pak10}. The blue curve in Fig.~\ref{fig3}\,a) shows that this ionization injection mechanism is very well localized.  Local injection is an indirect but striking consequence of the large dispersion effects: the intensity stays high only over a very short distance in the plasma, i.e. when the laser is fully compressed. Consequently, electrons are injected in the first cavity following the laser pulse, leading to a single electron bunch with duration $\sim 1\fs$, as illustrated in Fig.~\ref{fig3}\,b). The simulation reproduces the divergence ($\sim 20\mrad$), charge ($0.4\pC$) and energy distribution, see Fig.~\ref{fig3}\,c). The simulations also confirm that negative chirps yield no accelerated electrons and that a slight positive chirp optimizes the injection and acceleration process.
 
In addition, simulations suggest an explanation for the charge fluctuations observed in the experiment. Changing the CEP by $\pi/2$ causes the injected charge to increase by 15\%, while the energy distribution is not modified (not shown). As the CEP was not stabilized in our experiment, this accounts for some of the charge fluctuations. Simulations were also run by increasing the laser intensity by $3\%$, resulting in a charge increase of several 100's of fC. This indicates that the experiment operated close to the injection threshold and that using a slightly higher laser intensity might stabilize the injection process. 

This development will allow researchers to drive laser-plasma accelerators with more compact laser systems, thereby offering high repetition-rate operation and superior reliability. These femtosecond, kHz relativistic electron beams are now available and will open unique opportunities for a wide range of experiments such as femtosecond electron diffraction, femtosecond radiolysis or X-ray generation for fast scanning of dense objects.

\section*{Methods}
\begin{figure*}[t!]
\centerline{\includegraphics[width=1\textwidth]{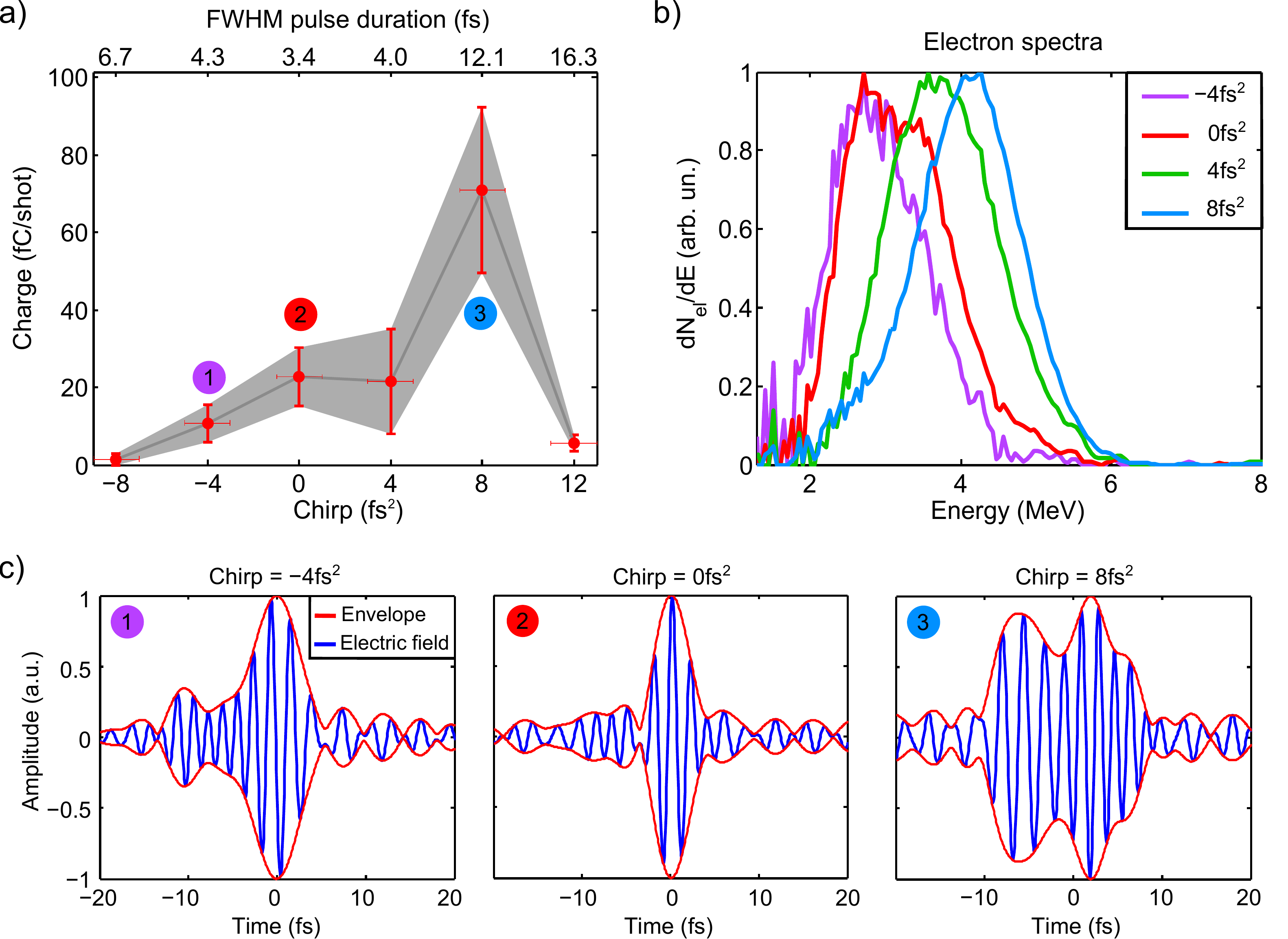}}
\caption{\textbf{Observation of dispersion effects.} Panel a): Evolution of accelerated charge with chirp of the pulse (in fs$^2$). The electron plasma density was $n_e=1.82\times10^{20}\cc$, corresponding to a capillary height of $90\mic$. Upper axis shows the estimated pulse FWHM duration. The grey area and vertical error bars represent the r.m.s. fluctuations over $20$ images (each of them averaged over $1$\,s\,$=1000$ shots). Horizontal error bars represent the uncertainty on the absolute value of the chirp. Panel b): Normalized electron spectra for different chirp values: $-4$\,fs$^2$ in purple, $0$\,fs$^2$ in red, $+4$\,fs$^2$ in green, $+8$\,fs$^2$ in blue. Panel c): Laser electric field (in blue) and envelope (in red) for three different chirp values.}\label{fig2}
\end{figure*}

\begin{figure*}[t!]
\centerline{\includegraphics[width=1\textwidth]{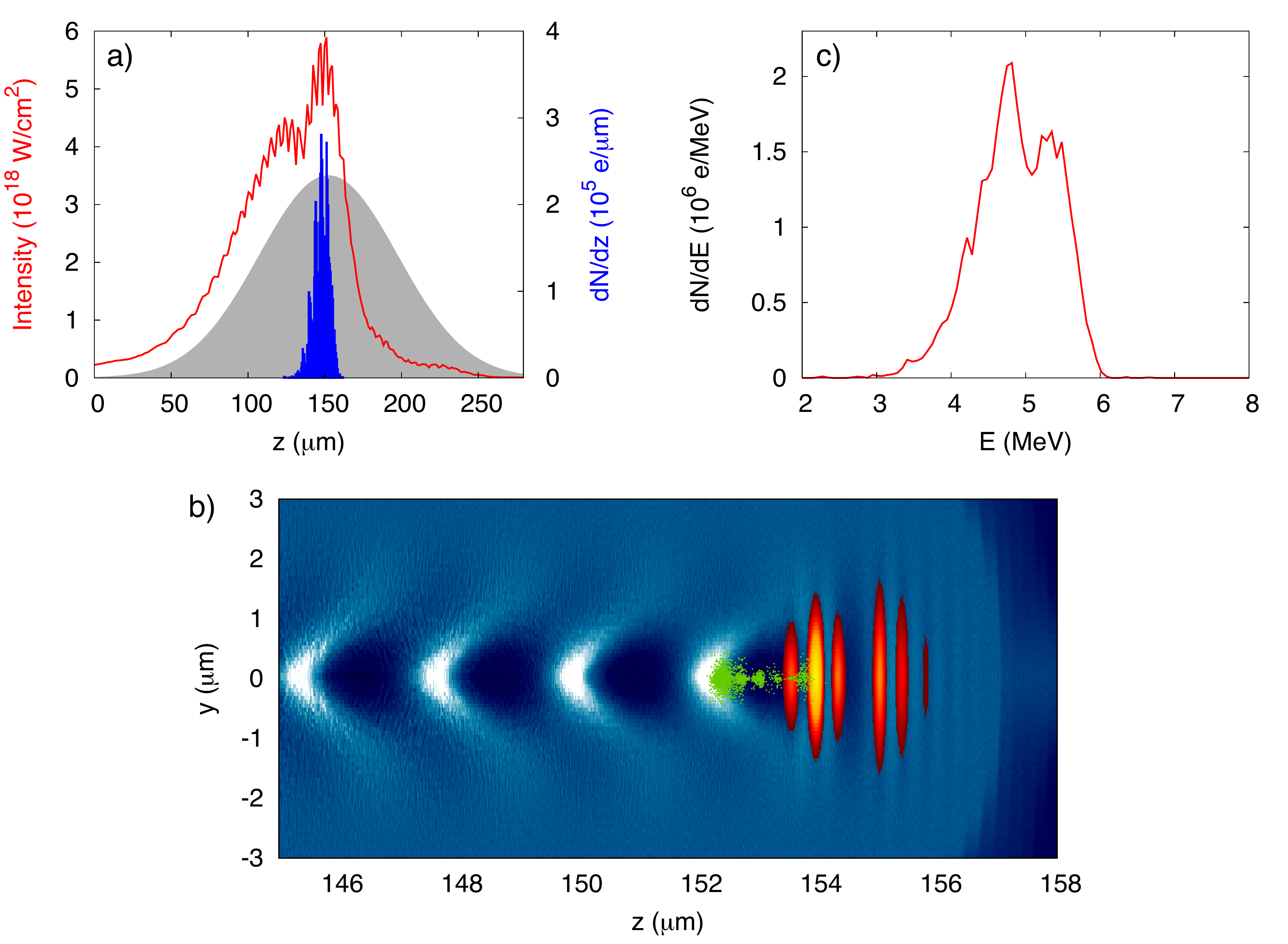}}
\caption{\textbf{Results of PIC simulations.} Panel a): Evolution of laser intensity (in red) and injected charge (in blue) during propagation in the plasma (density profile shown in grey) for a pulse with a $4\,\mathrm{fs}^2$ positive chirp. Panel b): Snapshot of the wakefield around the middle of the gas get. It shows the spatial distribution of electron density (in blue-white colorscale), the laser intensity (red-orange colorscale) and relativistic electrons (E>1.5 MeV) trapped in the wakefield (in green). Panel c): Electron energy spectrum at the accelerator exit.}\label{fig3}
\end{figure*}

\small{
\subsection{Single-cycle laser pulses}
The laser system is a double Chirped Pulse Amplification (CPA) system, delivering about 10~mJ in 25~fs at 800~nm wavelength with a temporal contrast better than $10^{10}$. The laser pulses are spectrally broadened in a 2.5~m long hollow core fiber filled with He gas. The pulses are post-compressed in vacuum to near-single-cycle pulses of 3.4~fs using a series of chirped mirrors. The beam is expanded in a reflective telescope to a transverse size of about 40~mm. The beam is then focused using an off-axis parabola with a 120~mm focal length, producing a high quality focal spot of $~3.5\mic$ at FWHM. The laser energy on target is typically in the range $2-2.5$~mJ. The laser pulse duration is measured directly in vacuum using the D-scan technique (Sphere Ultrafast Photonics \cite{mira12}): the laser pulse is frequency doubled in a thin BBO crystal and the second harmonic spectrum is measured for different insertions of two fused-silica wedges, providing a dispersion scan. An algorithm reconstructs the spectral amplitude and phase of the laser pulse, providing a complete temporal characterization of the laser pulse at best compression. The laser field inferred by such a measurement is shown in Fig.~\ref{fig2}c), in the middle panel for the unchirped case. Experimentally, the chirp is modified by changing the insertion of the two wedges in the laser beam: a positive chirp of $+4\,\mathrm{fs}^2$ is obtained by adding $100\,\mu\mathrm{m}$ of Fused-Silica.  For the chirped pulses, the temporal envelope is estimated by taking into account the spectral phase corresponding to this additional material. We model the spectral phase corresponding to the propagation in the wedges using Sellmeier's equation for the refractive index of Fused-Silica. This spectral phase is added to the measured phase of the unchirped laser field. Applying a Fourier transform, we then obtain the chirped laser field in the temporal domain, see left and right panels in Fig.~\ref{fig2}c). Finally, to obtain realistic values of the laser intensity, we take into account the real temporal and transverse distribution of the laser intensity, giving an estimated peak intensity of $I\simeq3\times10^{18}\wcm$. We also estimate that the intensity fluctuations, excluding possible fluctuations of the temporal envelope, are about $1\%$ r.m.s.

\subsection{Laser-plasma accelerator}
The laser beam is focused into a continuously flowing Nitrogen ($\mathrm{N}_2$) gas jet. The nozzle is a simple $100\mic$ diameter glass capillary that provides a sonic gas flow. We used a quadriwave lateral shearing interferometer (SID4 HR by PHASICS) to characterize the gas jet off-line, before and after experiments. The measurements show that the molecular density quickly drops above the capillary exit opening. We operate the experiment by sending the laser as close as $80\mic$ above the capillary exit, going closer would start to damage the capillary. At this position, the molecular profile can be approximated by a gaussian shape of $140\mic$ at FWHM. By applying a backing pressure of 20 bar, molecular densities up to $2\times10^{19}\cc$ can be obtained at the peak of the jet. The laser creates the plasma by barrier-suppression ionization of the Nitrogen atoms, up to $\mathrm{N^{5+}}$, providing electronic density up to $n_e=2\times10^{20}\cc$. Experimentally, the plasma density is modified by changing the backing pressure or by changing the respective height of the laser focus and the capillary exit. The density is limited by the gas load in the vacuum chamber: for a backing pressure of 20 bar of $\mathrm{N}_2$, the residual gas inside the chamber is below $10^{-2}$~mbar which is acceptable for running the experiment. Higher backing pressures make our turbomolecular pump fail, causing the background pressure to increase. Therefore, we cannot currently explore higher electron densities while operating at kHz repetition-rate, with a free-flowing gas jet.

Concerning electron detection, the electron beam profile is monitored using a CsI(Tl) phosphor screen imaged onto a 14-bit CCD camera. The electron energy distribution is obtained by sending the beam through a $500\mic$ lead pinhole followed by a pair of circular permanent magnets, providing a magnetic field peaking at $88$~mT. The phosphor screen was calibrated independently on a RF accelerator providing picosecond electron bunches at 3~MeV.
Finally, a small portion of the laser pulse was used as a probe in order to measure the electron density \textit{in situ} by transverse interferometry. A schematic of the experimental set-up is presented in the Suppl.Info.

The experiment was always run at kHz repetition-rate and all data and acquired images presented in this paper are averaged over 200 to 1000 shots. Statistics are usually presented by analyzing fluctuations over 20 acquisitions, meaning that each data point and error bars usually involve between 4000 and 20000 shots.

\subsection{PIC simulations}
Simulations were performed using CalderCirc \cite{lifs09}, a fully electromagnetic
3D code based on cylindrical coordinates $(r,z)$ and Fourier decomposition in the poloidal direction. The simulations were performed
using a mesh with $\Delta z=0.1\,k_0^{-1}$ and $\Delta
r=0.5\,k_0^{-1}$ (where $k_0=(2 \pi \lambda_0)^{-1}$ is the laser wave
vector and $\lambda_0=800$ nm), and the two first Fourier modes. The
neutral gas density profile was taken from
the experimental data. The simulations start with pure neutral Nitrogen, which
is ionized via tunnel ionization. The number of macro-particles per
cell before ionization is 500, which corresponds to 500$\times$5=2500
macro-electrons per cell in the region of full ionization of the
L-shell of Nitrogen. The temporal high frequency laser field for the different
values of chirp (-4 fs$^2$, 0 fs$^2$, 4 fs$^2$ and 8 fs$^2$) was
taken from experimental data (shown in figure \ref{fig2}c).

We explored the full range of the parameters spanned by the
experiment. Intensities were varied between $3 \times10^{18}\wcm$ and
$4 \times10^{18}\wcm$, and $\mathrm{N_2}$ densities from  $1.6
\times10^{19}\cc$ to  $2.5 \times10^{19}\cc$. For all cases, the
only injection mechanism found was ionization injection of electrons
coming from ionization of $\mathrm{N^{5+}}$, yielding a peaked electron
spectrum around 4-5 MeV. 

}

\section*{Acknowledgements}
We would like to acknowledge the help of the support team at the PHIL facility at LAL for the absolute calibration of our phosphor screens.
This work was funded by the European Research Council under Contract No. 306708, ERC Starting Grant FEMTOELEC. Financial support from the R\'egion Ile-de-France (under contract SESAME 2012-ATTOLITE), the Agence Nationale pour la Recherche (under contracts ANR-11-EQPX-005-ATTOLAB and ANR-14-CE32-0011-03 APERO) and ELI-HU Non-Profit Ltd (under contract NLO3.6LOA) is gratefully acknowledged. J.F. acknowledges fruitful discussions with Antoine Rousse, Victor Malka and Stephane Sebban.



\noindent
Correspondence and requests for materials should be addressed to: jerome.faure@ensta.fr,


\newpage

\end{document}